\documentclass[11pt,a4paper,twoside,groupcitations]{article}
\usepackage[T1]{fontenc}
\usepackage[ansinew]{inputenc}
\usepackage[english]{babel}
\usepackage{amsfonts}
\usepackage{amsmath}
\usepackage{bm}
\usepackage{array}
\usepackage{amsthm}
\usepackage{amssymb}
\usepackage{graphicx}
\usepackage{subfigure}
\usepackage{braket}
\usepackage{eucal}
\usepackage{verbatim}
\usepackage[table]{xcolor}
\usepackage{caption}
\usepackage{cite}
\usepackage{textcomp}
\usepackage{multicol}
\usepackage{tikz}
\usetikzlibrary{positioning,arrows}
\usetikzlibrary{decorations.pathmorphing}
\usetikzlibrary{decorations.markings}
\usetikzlibrary{calc,decorations.markings}
\usetikzlibrary{arrows,shapes}
\usetikzlibrary{matrix,arrows}
\usepackage{pgfplots}
\usepackage{xparse}
\definecolor{jade}{HTML}{00A86B}
\newcommand{\be}{\begin{eqnarray}}
\newcommand{\ee}{\end{eqnarray}}
\newcommand{\pro}[2]{\mbox{$\langle\, #1 \mid #2\,\rangle$}}

\renewcommand{\d}{\mbox{${\rm d}$}} %d differenziale non corsivo in math mode
\newcommand{\lp}{\ell_{\rm p}}
\newcommand{\mpl}{m_{\rm p}}
\newcommand{\gn}{G_{\rm N}}

\newcommand{\Rh}{R_{\rm H}}

\title{\bf Quantum dust cores of black holes}
\author{Roberto~Casadio$^{ab}$\thanks{E-mail: casadio@bo.infn.it}
\\
\\
$^a${\em Dipartimento di Fisica e Astronomia, Universit\`a di Bologna}
\\
{\em via Irnerio~46, 40126 Bologna, Italy}
\\
\\
$^b${\em I.N.F.N., Sezione di Bologna, I.S.~FLAG}
\\
{\em viale B.~Pichat~6/2, 40127 Bologna, Italy}}
\begin{document}
\maketitle
\begin{abstract}
We describe the ground state for a gravitationally collapsed ball of dust as the direct product of wavefunctions
for dust particles distributed over an arbitrary number of nested layers.
This allows us to estimate the expectation value of the global radius as well as the effective energy density
and pressures for the dust core of quantum black holes.
In particular, the size of the quantum core does not depend on the number of layers and the mass function
is shown to grow linearly with the areal radius up to the outermost layer.
\end{abstract}
\section{Introduction}
\setcounter{equation}{0}
\label{S:intro}
The singularities of known black hole solutions of the Einstein equations~\cite{HE} can be removed by imposing
regularity conditions on the (effective) energy density and scalar invariants inspired by classical
physics~\cite{Carballo-Rubio:2023mvr}.
This procedure usually induces the appearance (or fails to remove) an inner Cauchy horizon. 
A different framework, invoked for example in Ref.~\cite{Casadio:2023iqt}, can be implemented based on the possibility
that black hole interiors and the collapsed matter therein are described more accurately by quantum physics.
One can then consider an effective energy density $\rho\sim|\psi|^2$, where $\psi=\psi(r)$ is the wavefunction
of the fully collapsed matter source, such that the Misner-Sharp-Hernandez mass
function~\cite{Misner:1964je,Hernandez:1966zia} satisfies
\be
m(r)
\equiv
4\,\pi
\int_0^r
\rho(x)\,x^2\,\d x
\sim
4\,\pi
\int_0^r
|\psi(x)|^2\,x^2\,\d x
<
\infty
\qquad
{\rm for}
\ r>0
\ .
\label{Qcond}
\ee
This accommodates for $\rho\sim r^{-2}$ and $m\sim r$, which ensures that $m(0)=0$
and replaces the central singularity with an integrable one~\cite{Casadio:2021eio,Casadio:2022ndh}, that is a region
where the curvature invariants and the effective energy-momentum tensor diverge but their volume integrals
remain finite~\cite{Lukash:2013ts}.
\par
An explicit realisation for the inner core of a quantum black hole based on the Oppenheimer-Snyder model
of dust collapse~\cite{OS} was analysed in Refs.~\cite{Casadio:2021cbv,Casadio:2022pla,Casadio:2022epj}, where
only the outermost layer of dust was explicitly considered in an effective one-body approach.
This does not allow to estimate uniquely the size of the core or to obtain an effective energy density inside the
completely collapsed core itself, which are the main objectives of the present work.
To this end, we will here describe the ball as a sequence of layers~\cite{Tolman:1934za} containing dust particles,
whose trajectories are individually quantised as in Ref.~\cite{Casadio:2021cbv}.
A condition is then imposed to ensure that the fuzzy quantum layers defined by the positions of 
these particles remain orderly nested in the global quantum ground state.
This approach will confirm the expected quantum behaviour in Eq.~\eqref{Qcond} for the effective energy density
and mass function.
\par
It is important to stress that the above procedure differs from the canonical quantisation of
the Oppenheimer-Snyder model employed, for example,
in Refs.~\cite{Casadio:1998ta,Vaz:2011zz,Kiefer:2019csi,Husain:2022gwp,Giesel:2022rxi},
in which one starts from a reduced Einstein-Hilbert action for the areal radius of the ball.
Instead, we here quantise the trajectories of dust particles, which of course follow geodesics in the
classical theory, as more physically relevant degrees of freedom of the system,
similarly to what is done in the quantum mechanical description of the hydrogen atom.
We will then find that there exist ground states for the dust particles in each layer,
and a collective ground state for the whole core will then be built self-consistently,
starting from the quantum ground states of single dust particles.
We remark that no dynamical process will be analysed here which could possibly lead
to the formation of such a collective ground state, or of other quantum effects, like the Hawking evaporation.
\par
We will introduce the dynamical equation for dust particles in layers of the dust ball and derive
the single layer quantum states in the next Section; 
the global ground state is then constructed in Section~\ref{S:ground}, where its main features are also analysed;
concluding remarks and outlook will be given in Section~\ref{S:conc}.
\section{Quantum dust in a ball}
\setcounter{equation}{0}
\label{S:specttum}
Let us consider a perfectly isotropic ball of dust with total ADM~\cite{ADM} mass $M$ and areal radius $R=R(\tau)$,
where $\tau$ is the proper time measured by a clock comoving with the dust. 
Dust particles, which we assume have the same proper mass $\mu\ll M$, inside this collapsing ball will follow radial geodesics
$r=r(\tau)$ in the Schwarzschild spacetime metric~\footnote{We shall always use units with $c=1$ and often write the
Planck constant $\hbar=\lp\,\mpl$ and the Newton constant $\gn=\lp/\mpl$, where $\lp$ and $\mpl$ are the Planck length
and mass, respectively.}
\be
\d s^{2}
=
-\left(1-\frac{2\,\gn\,m}{r}\right)
\d t^{2}
+\left(1-\frac{2\,\gn\,m}{r}\right)^{-1}
\d r^{2}
+r^{2}\,\d\Omega^2
\ ,
\label{schw}
\ee
where $m=m(r)$ is the (constant Misner-Sharp-Hernandez) fraction of ADM mass inside the sphere of radius $r=r(\tau)$ and
$\d\Omega^2=\d\theta^{2}+\sin^{2}\theta\, \d\phi^{2}$.
Irrespectively of the mass profile $m=m(r)$, the classical dynamics predicts that an event horizon forms
when the surface areal radius $R(\tau)=2\,\gn\,M\equiv\Rh$ and the collapse will further proceed towards a singularity
in a finite proper time.
\par
We can discretise this ball by considering a spherical core of mass $\mu_0=\nu_0\,\mu=\epsilon_0\,M$
and radius $r=R_1(\tau)$ surrounded by $N$ comoving layers of inner radius $r=R_i(\tau)$, thickness $\Delta R_i=R_{i+1}-R_i$,
and mass $\mu_i=\epsilon_i\,M$, where $\epsilon_i$ is the fraction of ADM mass carried by the $\nu_i=\mu_i/\mu$
dust particles in the $i^{\rm th}$ layer.
The gravitational mass in the ball $r<R_i$ will be denoted by 
\be
M_i
=
\sum_{j=0}^{i-1}
\mu_j
=
M\,\sum_{j=0}^{i-1}\epsilon_j
=
\mu\,\sum_{i=0}^{i-1}\nu_j
\ ,
\ee
with $M_1=\mu_0$ and $M_{N+1}=M$. 
We also note that the radius $R_1$ and mass $M_1=\mu_0$ of the innermost core,~\footnote{The radius
$R_1$ can be interpreted as the size of the innermost core or the inner radius of the first layer around it.}
as well as the thickness $\Delta R_i$ of each layer, can be made arbitrarily small by increasing the number $N$ of layers
in the classical picture.
\par
The evolution of each layer can be derived by noting that dust particles located on the sphere of radius $r=R_i(\tau)$
will follow the radial geodesic equation
\be
\label{geod-part}
H_i
\equiv
\frac{P_i^{2}}{2\,\mu}
-\frac{\gn\,\mu\,M_i}{R_i}
=
\frac{\mu}{2}\left(\frac{E_i^{2}}{\mu^2}-1\right)
\equiv
\mathcal E_i
\ ,
\ee
where $P_i=\mu\,\d R_i/\d\tau$ is the radial momentum conjugated to $R=R_i(\tau)$, $E_i$ the conserved momentum
conjugated to $t=t_i(\tau)$ and the angular momentum conjugated to $\phi=\phi_i(\tau)$ was of course set to zero for
dust particles in a non-spinning ball~\cite{Casadio:2022epj}.
Notice that Eq.~\eqref{geod-part} depends on the (classically arbitrary) distribution of dust among the layers
of mass $\mu_{i>1}=M_{i+1}-M_i$ and the innermost spherical core of mass $\mu_0=M_1$.
This is the kind of improvement over previous works that we need in order to estimate the core profile
in the quantum ground state.~\footnote{The effective one-body approach in Ref.~\cite{Casadio:2021cbv}
is obtained by assuming $\mu\sim M$, which introduces undetermined numerical coefficients~\cite{Casadio:2022pla}
but leaves the final results qualitatively unaltered, as we shall duly comment in the following.} 
\par
With the canonical quantization prescription $P_i\mapsto\hat{P}_i=-i\,\hbar\,\partial_{R_i}$,
Eq.~\eqref{geod-part} becomes the time-independent Schr\"odinger equation
\be
\hat{H}_i\,\psi_{n_i}
=
\left[
-
\frac{\hbar^{2}}{2\,\mu}
\left(
\frac{\d^2}{\d R_i^2}
+
\frac{2}{R_i}\,
\frac{\d}{\d R_i}
\right)
-
\frac{\gn\,\mu\,M_i}{R_i}
\right]
\psi_{n_i}
=
\mathcal E_{n_i}\,
\psi_{n_i}
\ .
\ee
The above is formally the same as the equation for $s$-states of the hydrogen atom,
so that one can read out a Bohr radius
\be
a_i
=
\frac{\lp\,\mpl^2}{\mu\,M_i}
\ee
and the solutions are given by the Hamiltonian eigenfunctions
\be
\psi_{n_i}(R_i)
=
\sqrt{\frac{\mu^6\,M_i^{3}}{\pi\,\lp^{3}\,\mpl^{9}\,n_i^{5}}}\,
\exp\!\left(-\frac{\mu^2\,M_i\,R_i}{n_i\,\mpl^{3}\,\lp}\right)
L_{n_i-1}^{1}\!\!
\left(\frac{2\,\mu^2\,M_i\,R_i}{n_i\,\mpl^{3}\,\lp}\right)
\ ,
\label{radial-wavefunction}
\ee
where $L_{n-1}^1$ are Laguerre polynomials and ${n}_i=1,2\,\ldots$, corresponding to the eigenvalues 
\be
\mathcal E_{n_i}
=
-
\frac{\mu^3\,M_i^2}{2\,\mpl^4\,n_i^2}
\ .
\ee
The wavefunctions~\eqref{radial-wavefunction} are normalised in the scalar product which makes $\hat H_i$ Hermitian,
that is
\be
\pro{n_i}{n'_i}
=
4\,\pi
\int_0^\infty
R_i^2\,\psi_{n_i}^*(R_i)\,\psi_{n'_i}(R_i)\,
\d R_i
=
\delta_{n_i n'_i}
\ .
\ee
The expectation value of the areal radius on these eigenstates is given by
\be
\bar R_{n_i}
\equiv
\bra{n_i} \hat R_i \ket{n_i}
=
\frac{3\,\mpl^3\,\lp\,n_i^2}{2\,\mu^2\,M_i}
\ ,
\ee
with relative uncertainty
\be
\frac{\overline{\Delta R}_{n_i}}{\bar R_{n_i}}
\equiv
\frac{\sqrt{\bra{n_i}\hat R^2_i\ket{n_i}-\bar R_{n_i}^2}}{\bar R_{n_i}}
=
\frac{\sqrt{n_i^2+2}}{3\,n_i}
\ ,
\label{DR/R}
\ee
which approaches the minimum $\overline{\Delta R}_{n_i}\simeq \bar R_{n_i}/3$ for $n_i\gg 1$.
\par
By assuming that the conserved quantity $E_i$ remains well-defined for all the dust particles
in the allowed quantum states, we obtain the fundamental condition~\cite{Casadio:2021cbv}
\be
0
\le
\frac{E_i^2}{\mu^2}
=
1
+
\frac{2\,\mathcal E_i}{\mu}
=
1
-
\frac{\mu^2\,M_i^2}{\mpl^4\,n_i^2}
\ ,
\label{Emu}
\ee
which yields the lower bound for the single particle principal quantum numbers
\be
n_i
\ge
N_i
\equiv
\frac{\mu\,M_i}{\mpl^2}
\ .
\label{N_M}
\ee
Upon saturating the above bound, one then finds
\be
\bar R_{N_i}
=
\frac{3}{2}\,\gn\,M_i
\ ,
\label{RNi}
\ee
and the wavefunction for the $\nu_i$ particles in each layer is given by the same ground state
\be
\psi_{N_i}(R_i)
=
\sqrt{\frac{\mu\,\mpl}{\pi\,\lp^{3}\,M_i^{2}}}\,
\exp\!\left(-\frac{\mu\,R_i}{\mpl\,\lp}\right)
L_{\frac{\mu\,M_i}{\mpl^2}-1}^{1}\!\!
\left(\frac{2\,\mu\,R_i}{\mpl\,\lp}\right)
\ ,
\label{groundI}
\ee
where the values of $M_i$, hence $N_i$ in Eq.~\eqref{N_M}, must be such that $\bar R_i\lesssim \bar R_{i+1}$.
\par
From the above wavefunction, one can in principle determine the effective energy density inside each
layer as
\be
\rho_i
=
\mu\,\nu_i\,|\psi_{N_i}(r)|^2
%\simeq
%\mu\,\nu_i\,|\psi_{N_i}(\bar R_{N_i})|^2
\simeq
\mu\,\nu_i\,|\psi_{N_i}(3\,\gn\,M_i/2)|^2
\ ,
\ee
in which we approximated $r\simeq \bar R_{N_i}$ and used Eq.~\eqref{RNi}.
Clearly, the above expression depends on the number $\nu_i$ of dust particles in the $i^{\rm th}$ layer and the number
$\sum_{j=0}^i \nu_j$ of dust particles in the mass $M_i$ (and $N_i$), which are yet to be determined.
We shall see in the next Section how to estimate the distribution of particles $\nu_i$ and the corresponding 
energy density self-consistently.
\section{Multilayered ground state}
\setcounter{equation}{0}
\label{S:ground}
Since dust only interacts gravitationally, we can assume that the Hilbert space for the complete ball of mass
$M$ is given by the direct product $\mathcal H=\otimes_{i=1}^N\left( \otimes_{k=1}^{\nu_i} \mathcal H_i\right)$
of bound eigenstates~\eqref{radial-wavefunction} for the $\nu_i$ dust particles in each layer with
$\sum_{i=0}^N\nu_i=M/\mu$.
\par
We are here interested in the global ground state given by the product~\footnote{For our purpose, we do not
assume any specific statistics for the dust particles.
We expect that Pauli's exclusion principle will affect the analysis for fermions.}
\be
\ket{\{\nu_1,N_1\},\ldots,\{\nu_N,N_N\}}
=
\bigotimes_{i=1}^{N}
\ket{N_i}^{\nu_i}
\label{ground}
\ee
of single layer ground states, each one containing $\nu_i$ dust particles in the state $\ket{N_i}$,
and which further satisfy $\bar R_i\lesssim \bar R_{i+1}$.
Given the uncertainty~\eqref{DR/R}, the minimum thickness of the $i^{\rm th}$ layer of inner radius $\bar R_i$
must be of the order of $\overline{\Delta R}_i$ and the finest layering of the dust ball compatible with this
quantum description is given by $\bar R_{i+1}\simeq \bar R_i+\overline{\Delta R}_i\gtrsim 4\,\bar R_i/3$. 
On assuming $N_i\gg 1$ for all $i=1,\ldots,N$, one finds
\be
2\,\gn\,M_i
=
\frac{4}{3}\,\bar R_{N_i}
\simeq
\bar R_{N_i}
+
\overline{\Delta R}_{N_i}
\simeq
\bar R_{N_{i+1}}
=
\frac{3}{2}\,\gn\,M_{i+1}
\ ,
\ee
or $M_{i+1}\simeq 4\,M_{i}/3$, which implies that the mass of each layer $\mu_i\simeq M_i/3$.
The quantum numbers for the relevant single particle ground states in Eq.~\eqref{ground}
are therefore given by
\be
N_i
\simeq
\left(\frac{3}{4}\right)^{N-i+1}
\frac{\mu\,M}{\mpl^2}
\ .
\label{Ni}
\ee
\par
In particular, the quantum number for dust particles in the ground state of the outermost layer (with $i=N$)
is given by
\be
N_{\rm s}
\equiv
N_N
\simeq
\frac{3\,\mu\,M}{4\,\mpl^2}
\ ,
\label{Ns}
\ee
which yields the global ball radius
\be
R_{\rm s}
\equiv
\bar R_{N_{\rm s}}
+\overline{\Delta R}_{N_{\rm s}}
\simeq
\frac{3}{2}\,\gn\,M
\ .
\label{Rs}
\ee
Since $R_{\rm s}<\Rh$, the ground state of the dust ball can indeed be the core of a black hole, like it was
found in Refs.~\cite{Casadio:2021cbv,Casadio:2022pla,Casadio:2022epj}.
It is remarkable that the radius~\eqref{Rs} does not depend on the number $N$ of layers,
or any other quantity, except $M$.
In fact, $N$ only determines how finely we describe the central region of the ball.
In particular, the innermost core has radius $\bar R_1\simeq (3/4)^N R_{\rm s}$ and mass $\mu_0=M_1\simeq (3/4)^{N+1}\,M$.
Furthermore, it is interesting to note that multiplying $N_{\rm s}$ by the total number of particles $M/\mu$
recovers the black hole area quantisation~\footnote{For a solar mass ball made
of neutrons, the quantum number $N_{\rm s}\simeq 10^{19}$ (corresponding to $N_{\rm G}\simeq 10^{76}$), which makes
it practically impossible to study the wavefunctions}~\eqref{groundI} for realistic cases.
\be
\frac{M}{\mu}\,N_{\rm s}
\equiv
N_{\rm G}
\sim
\frac{M^2}{\mpl^2}
\sim
\frac{\Rh^2}{\lp^2}
\ ,
\label{Ng}
\ee
which again agrees with the results for the dust ball described as a single quantum
object~\cite{Casadio:2021cbv,Casadio:2022pla,Casadio:2022epj}
and with the coherent state description of the effective metric~\cite{Casadio:2021eio}.
We remark that the numerical prefactor in Eq.~\eqref{Ng} is not the same of the
Bekenstein-Hawking entropy~\cite{bekenstein}, but $N_{\rm G}$ is an integer nonetheless,
which suggests that mass and horizon area are quantised.~\footnote{The connection
of Eq.~\eqref{Ng} with the configurational entropy of the single dust core was investigated in Ref.~\cite{Casadio:2022pla}
and the present case is work in progress.}
It is certainly intriguing that this scaling of the black hole mass $M$ appears in several different approaches 
to quantum collapse and black holes
(see also, e.g.~Refs.~\cite{Kaup:1968zz,Mukhanov:1986me,Dvali:2011aa}).
\begin{figure}[t]
\centering
\includegraphics[width=10cm]{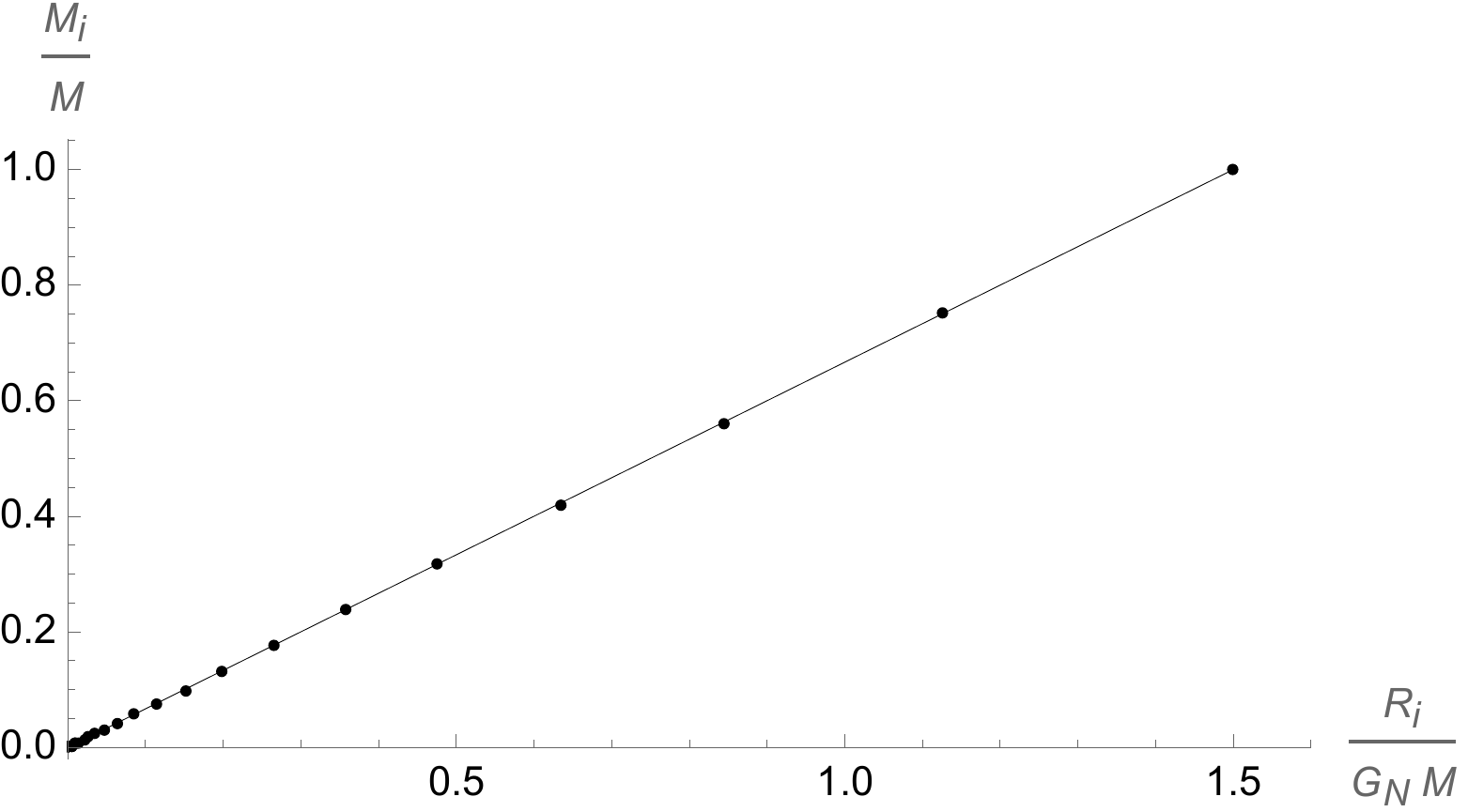}
\caption{Mass function $M_i$ (dots) for $N=100$ layers and its continuous approximation~\eqref{rho} (thin solid line).
The innermost core has radius $R_1\simeq 3\cdot 10^{-13}\,R_{\rm s}$ and mass $M_1\simeq 2\cdot 10^{-13}\,M$.}
\label{mR}
\end{figure}
\par
The crucial result for our present purpose is that the discrete mass function $M_i$ grows linearly with
the areal radius $R_i=\bar R_{N_i}$ in the collective ground state, irrespectively of the number of layers $N$
we employ to describe it.
One can therefore introduce a continuous effective energy density
\be
\rho
\simeq
\frac{M}{4\,\pi\,R_{\rm s}\,r^2}
\simeq
\frac{\mpl}{6\,\pi\,\lp\,r^2}
\ ,
\label{rho}
\ee
such that the Misner-Sharp-Hernandez mass function
\be
m(r)
=
4\,\pi
\int_0^r r^2\,\rho(r)\,\d r
=
\frac{2\,\mpl\,r}{3\,\lp}
\ee
equals the total ADM mass $M$ for $r=R_{\rm s}$ (see Fig.~\ref{mR}).
\par
Since dust particles in the ground state cannot collapse any further, the quantum core is necessarily
in equilibrium and one can determine the corresponding effective pressures from the isotropic metric
\be
\d s^2
&\!\!=\!\!&
-\left(1-\frac{2\,\gn\,m}{r}\right) \d t^2
+\left(1-\frac{2\,\gn\,m}{r}\right)^{-1} \d r^2
+
r^2\,\d\Omega^2
\nonumber
\\
&\!\!\simeq\!\!&
\frac{\d t^2}{3}\,
-3\, \d r^2
+
r^2\,\d\Omega^2
\ ,
\label{geff}
\ee
for $0\le r\le R_{\rm s}$.
From Eq.~\eqref{geff}, it is clear that there is no inner horizon inside the ground state core,
in agreement with the general results for spherical symmetry presented in Ref.~\cite{Casadio:2023iqt}.
We should further remark that the effective metric~\eqref{geff} cannot be used to describe any meaningful motion
inside the core, since matter is fully collapsed and cannot further evolve (except for the Hawking evaporation which
we neglect here).
The usual analysis of geodesics and geometric invariants therefore remains of purely formal value, 
as is perhaps the notion of Lorentzian signature inside the quantum core~\cite{Chen:2022eim}.
Nonetheless, the Ricci and Kretschmann scalars are given by 
$\mathcal R^2\simeq{\mathcal R}_{\alpha\beta\gamma\delta}\,{\mathcal R}^{\alpha\beta\gamma\delta}\simeq
{64}/{9\,r^4}$, whose square roots are integrable as anticipated in the Introduction.
\par
From the Einstein tensor of the metric~\eqref{geff}, one readily obtains the effective radial pressure 
\be
p_r
\simeq
-\frac{m'}{4\,\pi\,r^2}
\simeq
-\rho
\ ,
\ee
similar to other constructions for the black hole
interior (see~\cite{Carballo-Rubio:2023mvr,Brustein:2021lnr,Yokokura:2022kmq,Casadio:2020ueb,Farrah:2023opk}
and references therein),
and the tangential pressure (or tension) 
\be
p_\perp
\simeq
-\frac{m''}{8\,\pi\,r}
\simeq 0
\ ,
\ee
where primes denote derivatives with respect to $r$.
The vanishing of the tension inside each layer suggests that the system should be made to (differentially)
rotate easily~\cite{Casadio:2022epj} and is therefore likely unstable under perturbations of the angular momentum.
\begin{figure}[t]
\centering
\includegraphics[width=10cm]{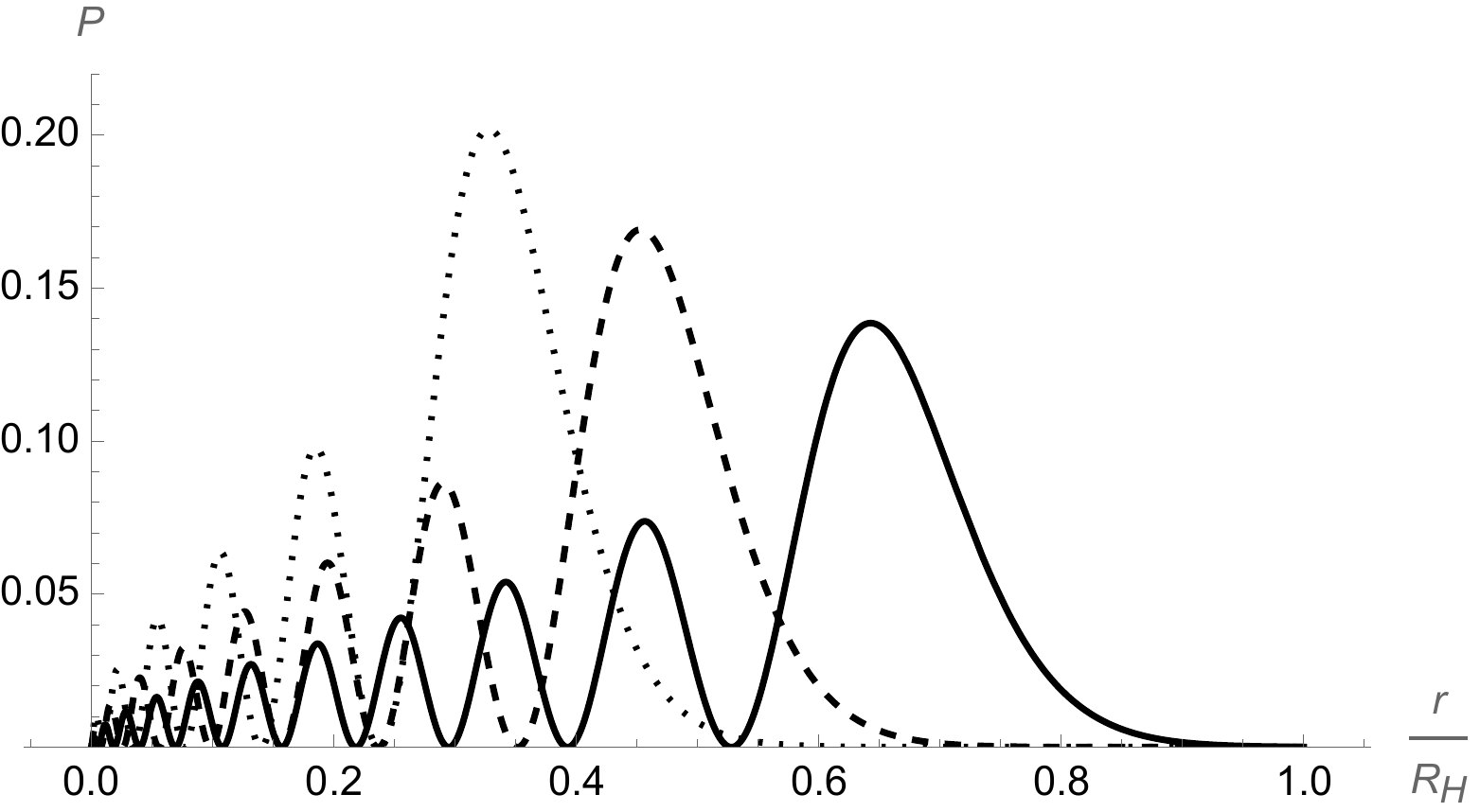}
\caption{Probability density~\eqref{Pr} for $N=3$, $\mu=\mpl/10$, $M_3=1100\,\mu$, corresponding
to a black hole with $M\simeq 150\,\mpl$ and $\Rh\simeq 300\,\lp$.}
\label{3layers}
\end{figure}
\par
Note that the outermost layer has an estimated thickness $\overline{\Delta R}_N\simeq R_{\rm s}/4\simeq 3\,\gn\,M/8$
and contains $\mu_N/M\simeq 1/4$ of the total mass.
A more accurate description near the surface of the core that matches smoothly with the outer Schwarzschild
geometry of ADM mass $M$ can then be obtained from the effective energy density 
\be
\rho
\simeq
\frac{M}{4}\,|\psi_{N_{\rm s}}(r)|^2
\ .
\ee
The mass function is therefore not expected to remain linear for $r\gtrsim \bar R_N$ and the tension will not vanish
near the surface of the core.
To clarify this point, we plot the probability densities 
\be
\mathcal{P}_i
=
4\,\pi\,r^2\,|\psi_{N_i}(r)|^2
=
4\,\pi\,\mu_i^{-1}\,\rho_i(r)
\label{Pr}
\ee
for an example with $N=3$ layers in Fig.~\ref{3layers}, from which it appears that the probability of finding 
a particle of the $i^{\rm th}$ layer inside both narrower and broader layers with $j\not=i$ 
is not zero.
In particular, the wavefunction of dust particles in the $i^{\rm th}$ layer overlaps with those in
all the layers $j<i$.
Since this fact was neglected in the derivation of the continuous approximation~\eqref{rho} from the 
discrete mass function $M_i$, we expect that the actual density decreases somewhat faster from the centre,
which could particularly affect the amount of dust in the outermost layer.
A more accurate description of the effective energy density of the outermost layer is left
for future investigations.
\section{Conclusions and outlook}
\setcounter{equation}{0}
\label{S:conc}
We have improved on the quantum (one-body) description of the ball of dust introduced in Ref.~\cite{Casadio:2021cbv} 
by dividing the ball into $N$ layers, each of which contains dust particles described by quantum states of
the same general relativistic dynamics. 
By requiring that the thickness of each layer be given by the quantum uncertainty in the location of the particles therein,
we have obtained a unique collective ground state whereby the radius of the ball is determined by the total ADM mass $M$, 
irrespectively of $N$, and the horizon area is quantised according to Eq.~\eqref{Ng},
in qualitative agreement with Bekenstein's conjecture of the black hole area quantisation~\cite{bekenstein}.
We should remark that these features follow from assuming that the number of dust particles in each layer is large and
departures are expected to occur when this condition is violated (that is, for small black holes).
\par
A continuous effective energy density was also estimated to match the discrete mass distribution found for $N$ layers,
which turns out to be precisely of the form in Eq.~\eqref{Qcond} almost everywhere inside the core. 
This latter result lends further support to the picture of quantum black holes described in Ref.~\cite{Casadio:2023iqt}.
An effective radial pressure of quantum origin exactly opposite the energy density sustains the ground state,
whereas the tangential pressure is found to vanish, again almost everywhere inside the core, thus suggesting that
the nature of dust is unaffected by quantum gravity in a perfectly spherical configuration. 
\par
Of course, the present work is not free from shortcomings and limiting assumptions.
As we mentioned in the Introduction, we {\em a priori\/} considered static configurations for the dust particles
and did not even attempt at analysing the time evolution that could lead to the formation of the collective ground state.
In principle, such an evolution should occur as dust particles progressively jump from higher excited states to lower
levels~\cite{Casadio:2022pla}, a clearly very complex process given the huge number of particles in an astrophysical object. 
Furthermore, we considered dust particles when one would eventually like to describe matter by means of quantum
excitations of standard model fields and all of their interactions.
The necessary existence of other interactions will give rise to additional pressure terms and could very significantly
influence both the global size of the core and the effective energy density.
Even without the inclusion of pressure terms, we argued that the energy density $\rho\sim |\psi_{N_{\rm s}}|^2$
in the outermost layer implies a different behaviour for the mass function
and a non-vanishing tension at the surface of the core.
A more accurate description of the core surface will become particularly relevant to understanding
what happens when more matter accretes or the Hawking effect evaporates the core.
We leave all of these complex issues for future investigations. 
\section*{Acknowledgments}
R.C.~is partially supported by the INFN grant FLAG and his work has also been carried out in
the framework of activities of the National Group of Mathematical Physics (GNFM, INdAM).

\begin{thebibliography}{99}
%
%
\bibitem{HE}
S.~W.~Hawking and G.~F.~R.~Ellis,
``The Large Scale Structure of Space-Time,''
%  doi:10.1017/CBO9780511524646
(Cambridge University Press, Cambridge, 1973)
%%CITATION = doi:10.1017/CBO9780511524646;%%
%1004 citations counted in INSPIRE as of 25 Oct 2018
%
%\cite{Carballo-Rubio:2023mvr}
\bibitem{Carballo-Rubio:2023mvr}
R.~Carballo-Rubio, F.~Di Filippo, S.~Liberati and M.~Visser,
``Singularity-free gravitational collapse: From regular black holes to horizonless objects,''
[arXiv:2302.00028 [gr-qc]].
%0 citations counted in INSPIRE as of 05 Mar 2023%
%
%\cite{Casadio:2023iqt}
\bibitem{Casadio:2023iqt}
R.~Casadio, A.~Giusti and J.~Ovalle,
%``Quantum rotating black holes,''
JHEP \textbf{05} (2023) 118
%doi:10.1007/JHEP05(2023)118
[arXiv:2303.02713 [gr-qc]].
%6 citations counted in INSPIRE as of 05 Jun 2023
% 
%\cite{Misner:1964je}
\bibitem{Misner:1964je}
C.~W.~Misner and D.~H.~Sharp,
%``Relativistic equations for adiabatic, spherically symmetric gravitational collapse,''
Phys. Rev. \textbf{136} (1964), B571.
%doi:10.1103/PhysRev.136.B571
%940 citations counted in INSPIRE as of 13 Jan 2023
%
%\cite{Hernandez:1966zia}
\bibitem{Hernandez:1966zia}
W.~C.~Hernandez and C.~W.~Misner,
%``Observer Time as a Coordinate in Relativistic Spherical Hydrodynamics,''
Astrophys. J. \textbf{143} (1966) 452.
%doi:10.1086/148525
%160 citations counted in INSPIRE as of 13 Jan 2023
%
%\cite{Casadio:2021eio}
\bibitem{Casadio:2021eio}
R.~Casadio,
%``Geometry and thermodynamics of coherent quantum black holes,''
Int. J. Mod. Phys. D \textbf{31} (2022) 2250128
%doi:10.1142/S0218271822501280
[arXiv:2103.00183 [gr-qc]].
%10 citations counted in INSPIRE as of 13 Jan 2023
%
%\cite{Casadio:2022ndh}
\bibitem{Casadio:2022ndh}
R.~Casadio, A.~Giusti and J.~Ovalle,
%``Quantum Reissner-Nordstr\"om geometry: Singularity and Cauchy horizon,''
Phys. Rev. D \textbf{105} (2022) 124026
%doi:10.1103/PhysRevD.105.124026
[arXiv:2203.03252 [gr-qc]].
%4 citations counted in INSPIRE as of 14 Oct 2022
%
%\cite{Lukash:2013ts}
\bibitem{Lukash:2013ts}
V.~N.~Lukash and V.~N.~Strokov,
%``Space-Times with Integrable Singularity,''
Int. J. Mod. Phys. A \textbf{28} (2013) 1350007
%doi:10.1142/S0217751X13500073
[arXiv:1301.5544 [gr-qc]].
%13 citations counted in INSPIRE as of 05 Mar 2023
%
%\cite{Oppenheimer:1939ue}
\bibitem{OS}
J.~R.~Oppenheimer and H.~Snyder,
%``On Continued gravitational contraction,''
Phys. Rev. \textbf{56} (1939) 455.
%doi:10.1103/PhysRev.56.455
%907 citations counted in INSPIRE as of 25 Mar 2021
%
%\cite{Casadio:2021cbv}
\bibitem{Casadio:2021cbv}
R.~Casadio,
%``A quantum bound on the compactness,''
Eur. Phys. J. C \textbf{82} (2022) 10
%doi:10.1140/epjc/s10052-021-09980-2
[arXiv:2103.14582 [gr-qc]].
%8 citations counted in INSPIRE as of 18 Feb 2022
%
%\cite{Casadio:2022pla}
\bibitem{Casadio:2022pla}
R.~Casadio, R.~da Rocha, P.~Meert, L.~Tabarroni and W.~Barreto,
%``Configurational entropy of black hole quantum cores,''
Class. Quant. Grav. \textbf{40} (2023) 075014
%doi:10.1088/1361-6382/acbe89
[arXiv:2206.10398 [gr-qc]].
%4 citations counted in INSPIRE as of 09 Apr 2023
%
%\cite{Casadio:2022epj}
\bibitem{Casadio:2022epj}
R.~Casadio and L.~Tabarroni,
%``Slowly rotating quantum dust cores and black holes,''
Eur. Phys. J. Plus \textbf{138} (2023) 104
%doi:10.1140/epjp/s13360-023-03705-y
[arXiv:2212.05514 [gr-qc]].
%3 citations counted in INSPIRE as of 09 Apr 2023
%
%\cite{Tolman:1934za}
\bibitem{Tolman:1934za}
R.~C.~Tolman,
%``Effect of imhomogeneity on cosmological models,''
Proc. Nat. Acad. Sci. \textbf{20} (1934) 169.
%doi:10.1073/pnas.20.3.169
%805 citations counted in INSPIRE as of 11 Apr 2023
%
%\cite{Casadio:1998ta}
\bibitem{Casadio:1998ta}
R.~Casadio,
%``Hamiltonian formalism for the Oppenheimer-Snyder model,''
Phys. Rev. D \textbf{58} (1998) 064013
%doi:10.1103/PhysRevD.58.064013
[arXiv:gr-qc/9804021 [gr-qc]].
%8 citations counted in INSPIRE as of 02 Jun 2023
%
%\cite{Vaz:2011zz}
\bibitem{Vaz:2011zz}
C.~Vaz and L.~Witten,
%``Canonical quantization of spherically symmetric dust collapse,''
Gen. Rel. Grav. \textbf{43} (2011) 3429
%doi:10.1007/s10714-011-1240-4
[arXiv:1111.6821 [gr-qc]].
%9 citations counted in INSPIRE as of 02 Jun 2023
%
%\cite{Kiefer:2019csi}
\bibitem{Kiefer:2019csi}
C.~Kiefer and T.~Schmitz,
%``Singularity avoidance for collapsing quantum dust in the Lema\^\i{}tre-Tolman-Bondi model,''
Phys. Rev. D \textbf{99} (2019) 126010
%doi:10.1103/PhysRevD.99.126010
[arXiv:1904.13220 [gr-qc]].
%33 citations counted in INSPIRE as of 02 Jun 2023
%
%\cite{Husain:2022gwp}
\bibitem{Husain:2022gwp}
V.~Husain, J.~G.~Kelly, R.~Santacruz and E.~Wilson-Ewing,
%``Fate of quantum black holes,''
Phys. Rev. D \textbf{106} (2022) 024014
%doi:10.1103/PhysRevD.106.024014
[arXiv:2203.04238 [gr-qc]].
%14 citations counted in INSPIRE as of 02 Jun 2023
%
%\cite{Giesel:2022rxi}
\bibitem{Giesel:2022rxi}
K.~Giesel, M.~Han, B.~F.~Li, H.~Liu and P.~Singh,
%``Spherical symmetric gravitational collapse of a dust cloud: Polymerized dynamics in reduced phase space,''
Phys. Rev. D \textbf{107} (2023) 044047
%doi:10.1103/PhysRevD.107.044047
[arXiv:2212.01930 [gr-qc]].
%5 citations counted in INSPIRE as of 02 Jun 2023
%
\bibitem{ADM}
R.L.~Arnowitt, S.~Deser and C.W.~Misner,
%``Dynamical Structure and Definition of Energy in General Relativity,''
Phys.\ Rev.\  {\bf 116} (1959) 1322.
%  doi:10.1103/PhysRev.116.1322
%%CITATION = doi:10.1103/PhysRev.116.1322;%%
%270 citations counted in INSPIRE as of 19 Jan 2017
%
%\cite{Bekenstein:1973ur}
\bibitem{bekenstein}
J.~D.~Bekenstein,
%``Black holes and entropy,''
Phys. Rev. D \textbf{7} (1973) 2333.
%doi:10.1103/PhysRevD.7.2333
%5062 citations counted in INSPIRE as of 06 Apr 2021
%
%\cite{Chen:2022eim}
\bibitem{Chen:2022eim}
P.~Chen, M.~Sasaki, D.~h.~Yeom and J.~Yoon,
%``Resolving information loss paradox with Euclidean path integral,''
Int. J. Mod. Phys. D \textbf{31} (2022) 2242001
%doi:10.1142/S0218271822420019
[arXiv:2205.08320 [gr-qc]].
%1 citations counted in INSPIRE as of 13 Apr 2023
%
%\cite{Kaup:1968zz}
\bibitem{Kaup:1968zz}
D.~J.~Kaup,
%``Klein-Gordon Geon,''
Phys. Rev. \textbf{172} (1968) 1331
%doi:10.1103/PhysRev.172.1331
%609 citations counted in INSPIRE as of 03 Jun 2023
%
%\cite{Mukhanov:1986me}
\bibitem{Mukhanov:1986me}
V.~F.~Mukhanov,
%``ARE BLACK HOLES QUANTIZED?,''
JETP Lett. \textbf{44} (1986) 63.
%165 citations counted in INSPIRE as of 12 Apr 2023
%
%\cite{Dvali:2011aa}
\bibitem{Dvali:2011aa}
G.~Dvali and C.~Gomez,
%``Black Hole's Quantum N-Portrait,''
Fortsch. Phys. \textbf{61} (2013) 742
%doi:10.1002/prop.201300001
[arXiv:1112.3359 [hep-th]].
%365 citations counted in INSPIRE as of 03 Jun 2023
%
%\cite{Brustein:2021lnr}
\bibitem{Brustein:2021lnr}
R.~Brustein, A.~J.~M.~Medved and T.~Simhon,
%``Black holes as frozen stars,''
Phys. Rev. D \textbf{105} (2022) 024019
%doi:10.1103/PhysRevD.105.024019
[arXiv:2109.10017 [gr-qc]].
%3 citations counted in INSPIRE as of 12 Apr 2023
%
%\cite{Yokokura:2022kmq}
\bibitem{Yokokura:2022kmq}
Y.~Yokokura,
``Entropy-Area Law from Interior Semi-classical Degrees of Freedom,''
[arXiv:2207.14274 [hep-th]].
%1 citations counted in INSPIRE as of 11 Apr 2023
%
%\cite{Casadio:2020ueb}
\bibitem{Casadio:2020ueb}
R.~Casadio, M.~Lenzi and A.~Ciarfella,
%``Quantum black holes in bootstrapped Newtonian gravity,''
Phys. Rev. D \textbf{101} (2020) 124032
%doi:10.1103/PhysRevD.101.124032
[arXiv:2002.00221 [gr-qc]].
%16 citations counted in INSPIRE as of 12 Apr 2023
%
%\cite{Farrah:2023opk}
\bibitem{Farrah:2023opk}
D.~Farrah
%K.~S.~Croker, G.~Tarl\'e, V.~Faraoni, S.~Petty, J.~Afonso, N.~Fernandez, K.~A.~Nishimura, C.~Pearson and L.~Wang, 
\textit{et al.}
%``Observational Evidence for Cosmological Coupling of Black Holes and its Implications for an Astrophysical Source of Dark Energy,''
Astrophys. J. Lett. \textbf{944} (2023) L31
%doi:10.3847/2041-8213/acb704
[arXiv:2302.07878 [astro-ph.CO]].
%13 citations counted in INSPIRE as of 02 Jun 2023
%
\end{thebibliography}
\end{document}